\documentclass[sigconf,nonacm]{acmart}

\setcopyright{none}
\renewcommand\footnotetextcopyrightpermission[1]{}

\usepackage{booktabs}
\usepackage{multirow}
\usepackage{graphicx}
\usepackage{amsmath}
\usepackage{adjustbox}
\usepackage{placeins}
\usepackage{float}

\begin{document}

\title{Multilingual Humour-Aware Retrieval with Dense and Re-Ranking Models}

\author{Georgios Arampatzis}
\authornote{Corresponding author.}
\affiliation{
  \institution{Democritus University of Thrace}
  \department{Department of Electrical and Computer Engineering}
  \city{Xanthi}
  \country{Greece}
}
\email{geoaramp@ee.duth.gr}

\author{Avi Arampatzis}
\affiliation{
  \institution{Democritus University of Thrace}
  \department{Department of Electrical and Computer Engineering}
  \city{Xanthi}
  \country{Greece}
}
\email{avi@ee.duth.gr}

\pagestyle{plain}
\thispagestyle{plain}

\begin{abstract}
Humour-aware information retrieval (IR) poses unique challenges beyond standard semantic retrieval, as systems must account not only for topical relevance but also for humour-specific linguistic phenomena such as wordplay, phonetic ambiguity, and polysemy. In this paper, Team DUTH studies multilingual humour-aware information retrieval using the CLEF 2025 JOKER Task~1 benchmark, which evaluates humour retrieval in English and Portuguese. Our approach combines multilingual XLM-RoBERTa-based dense retrieval with additional system variants, including neural re-ranking, in order to assess the extent to which general-purpose Transformer models can capture humour-specific relevance.

The results reveal substantial cross-lingual variation. While the Portuguese runs demonstrate comparatively strong performance across MAP, MRR, and early precision metrics, the English runs perform significantly worse, with relevant humorous documents frequently appearing at lower ranks. These findings highlight the limitations of purely semantic dense representations for humour retrieval, particularly when humour depends on surface-level cues that are not explicitly modelled by multilingual encoders. We further analyse contributing factors to this discrepancy, including dataset characteristics, query--document alignment, and variation in humour mechanisms.

Overall, the Team DUTH experiments establish multilingual dense-retrieval and re-ranking baselines and provide insights into the challenges of modelling humour-aware relevance within the JOKER framework.
\end{abstract}

\keywords{Information Retrieval, Humour Retrieval, Dense Retrieval, CLEF, JOKER Task}

\maketitle


\section{Introduction}

Humour-aware information retrieval (IR) has recently emerged as a challenging research problem at the intersection of information retrieval, computational humour, and multilingual natural language processing. Unlike traditional retrieval settings---where topical similarity is typically the primary criterion---humour retrieval requires systems to recognise complex linguistic mechanisms such as wordplay, puns, phonetic ambiguity, lexical polysemy, metaphorical structures, and contextual reinterpretation. These phenomena introduce additional layers of complexity, as humour often depends on subtle surface-level or phonological cues that are not captured by conventional semantic similarity models.

Early approaches to humour retrieval relied primarily on lexical similarity and collaborative filtering techniques, as demonstrated by systems such as Jester and early joke retrieval experiments~\cite{friedland2008joke, gupta1999jester}. In parallel, research in computational humour has expanded significantly, covering tasks such as humour classification, pun detection, pun interpretation, and humour translation (e.g.,~\cite{erkamillerpun2020, miller2022humour}). Over the past decade, Transformer-based models have become the dominant paradigm in multilingual semantic modelling and have achieved strong performance in general-purpose IR tasks, particularly through dense retrieval methods based on Sentence-BERT, MPNet, and multilingual RoBERTa~\cite{reimers2019sentencebert, song2020mpnet}. Nevertheless, these models remain limited in their ability to capture humour-specific linguistic structures, which often rely on irregular, non-compositional, or surface-level cues.

The CLEF JOKER initiative has played a central role in advancing and standardising the evaluation of computational humour. Its scope has evolved from automatic wordplay analysis and pun interpretation to humour translation and, more recently, humour-aware retrieval. Task~1 of the 2025 edition continues this trajectory by introducing multilingual IR benchmarks in English and European Portuguese, with relevance judgements based on both topical alignment and humour-specific equivalence~\cite{joker2025overview}. Notably, the dataset includes both human-written and machine-generated texts, introducing substantial stylistic variability and increasing the difficulty for semantic retrieval systems.

Team DUTH has extensive experience in multilingual NLP shared tasks using Transformer-based architectures. Previous work includes intimacy analysis at SemEval 2023~\cite{arampatzis2023semeval}, multilingual sexism detection with soft labels in EXIST~2025~\cite{arampatzis2025exist}, pun and proper-name translation in JOKER 2025 Tasks~2 and~3~\cite{arampatzis2025joker}, and scientific text simplification and hallucination detection in the SimpleText track~\cite{arampatzis2025simpletext}. This body of work provides a strong foundation for investigating humour-aware retrieval, particularly in multilingual settings where semantic and stylistic signals interact in complex ways.

Motivated by these developments, this study evaluates multilingual humour-aware retrieval using the JOKER 2025 Task~1 benchmark. We examine dense retrieval models based on XLM-RoBERTa together with additional system variants, including fine-tuned models and neural re-ranking. This setup allows us to analyse how well general-purpose multilingual representations, task-specific supervision, and interaction-based re-ranking capture humour-specific relevance. Our results reveal pronounced cross-lingual variation: while performance in Portuguese is comparatively stable and effective, performance in English remains substantially weaker, suggesting limited generalisation and an overreliance on semantic similarity alone.

The remainder of this paper is structured as follows. Section~2 reviews prior
work on computational humour and semantic retrieval. Section~3 describes the
JOKER Task~1 dataset and evaluation framework. Section~4 presents the
multilingual retrieval and re-ranking approaches. Section~5 outlines the
experimental setup. Section~6 reports and analyses retrieval performance in
English and Portuguese. Section~7 discusses the implications of the findings.
Section~8 presents the limitations of the study, and Section~9 concludes the
paper and highlights directions for future research.

\section{Related Work}
\label{sec:related}

Research in computational humour has expanded steadily over the past two
decades, encompassing tasks such as humour recognition, joke retrieval, pun
detection, pun interpretation, and humour translation. Early work focused
primarily on retrieval-style problems. The Jester~2.0 system explored
collaborative filtering for joke recommendation~\cite{gupta1999jester}, while
Friedland and Allan~\cite{friedland2008joke} investigated methods for
retrieving semantically similar jokes despite surface-level variation. These
studies demonstrated that humour retrieval is feasible but highly sensitive to
lexical variation, stylistic reformulation, and ambiguity.

A substantial body of research has since emerged on automatic wordplay
analysis. Miller and colleagues~\cite{miller2017pun}
formalised tasks such as pun detection and interpretation, introducing
annotation schemes and benchmark datasets that highlight the linguistic
complexity of humour. More recent work has extended these approaches to
multilingual and cross-lingual settings, including the translation of puns and
humour-bearing expressions, where semantic preservation must be balanced with
cultural and linguistic constraints. The JOKER tasks (2022--2025) have
contributed a series of benchmark datasets and evaluation frameworks for
wordplay classification, pun localisation, interpretation, humour translation,
and humour-aware retrieval~\cite{joker2025overview, joker2024overview,
joker2023overview, joker2023pun, joker2022overview}.

In parallel, the information retrieval community has made rapid progress in
semantic retrieval through the use of Transformer-based encoders. Dense vector
representations derived from models such as
Sentence-BERT~\cite{reimers2019sentencebert}, MPNet~\cite{song2020mpnet}, and
multilingual Transformer architectures have significantly improved semantic
matching across languages. However, these models are primarily optimised for
capturing topical similarity rather than the surface-level linguistic cues that
are critical for humour, including phonetic similarity, homophony,
morphological variation, and syntactic ambiguity. As a result, dense retrieval
models may fail in humour-centric settings, where relevance depends not only on
semantic alignment but also on shared humour mechanisms.

Hybrid and multi-stage retrieval architectures provide one way to address the
limitations of purely semantic matching. Lexical models such as BM25 remain
strong and interpretable baselines, while pseudo-relevance feedback and rank
fusion methods such as RM3 and Reciprocal Rank Fusion (RRF) can improve
robustness by combining complementary retrieval signals. These ideas have been
effective in recent DUTH retrieval systems for task-oriented product search and
memory-based retrieval, where calibrated RM3, lexical ensembles, and RRF were
used to bridge vocabulary mismatch and improve recall under ambiguous or
underspecified queries~\cite{arampatzis2025productsearch,
arampatzis2025tot}. Similarly, sparse--neural pipelines combining BM25 with
cross-encoder re-ranking and fusion have shown that lightweight re-rankers can
improve early ranking quality while preserving reproducibility and computational
efficiency~\cite{arampatzis2025ragtime}.

Retrieval-augmented and justification-oriented evaluation has also become
increasingly relevant for modern IR systems. Recent DUTH work on the TREC RAG
and DRAGUN tracks explored hybrid justification labelling, retrieval-grounded
reporting, and question generation for credibility assessment, highlighting the
importance of combining retrieval signals, semantic modelling, and transparent
evaluation pipelines~\cite{arampatzis2025rag, arampatzis2025dragun}. Although
these tasks differ from humour-aware retrieval, they share a common emphasis on
evidence selection, ranking quality, reproducibility, and the interaction
between lexical and semantic signals.

The CLEF JOKER Task~1 combines elements from both research directions: it is
formulated as an IR task, yet its relevance judgements are strongly conditioned
on humour-specific linguistic properties. While lexical retrieval captures
surface overlap and dense retrieval captures semantic relatedness, neither
signal alone is sufficient for humour-aware relevance when the decisive cue is a
pun, ambiguity, phonetic resemblance, or shared wordplay mechanism. This
motivates the evaluation of dense retrieval, fine-tuned variants, and
re-ranking approaches within a multilingual humour-aware retrieval framework.

Team DUTH has contributed to multiple multilingual NLP and IR shared tasks
using Transformer-based, lexical, and hybrid architectures. Previous work
includes intimacy analysis at SemEval 2023~\cite{arampatzis2023semeval},
multilingual sexism detection with soft labels at EXIST~2025
~\cite{arampatzis2025exist}, pun and proper-name translation in JOKER 2025
Tasks~2 and~3~\cite{arampatzis2025joker}, and scientific text simplification
and hallucination detection in the SimpleText track
~\cite{arampatzis2025simpletext}. Additional recent work has explored
retrieval-augmented reporting and question generation for news credibility,
hybrid justification labelling for RAG evaluation, lexical product search with
RM3 and fusion, sparse--neural passage re-ranking, and Tip-of-the-Tongue
retrieval~\cite{arampatzis2025dragun, arampatzis2025rag,
arampatzis2025productsearch, arampatzis2025ragtime, arampatzis2025tot}. These
efforts share a common emphasis on reproducibility, controlled experimentation,
and the interaction between lexical, semantic, and task-specific signals.

Despite these advances, humour-aware IR remains relatively underexplored. The
integration of semantic retrieval with humour modelling introduces unresolved
challenges related to cross-lingual transfer, the representation of
surface-level linguistic cues, and query--document alignment under humour
transformations. In this work, we contribute multilingual retrieval baselines,
including dense retrieval, fine-tuned variants, and neural re-ranking, and
provide insights into the limitations of semantic embeddings when applied to
humour-aware retrieval.

\section{Task Description}
\label{sec:task}

CLEF JOKER Task~1 addresses \textit{humour-aware information retrieval}, a setting in which systems must retrieve documents that are not only topically relevant to a given humorous query but also exhibit equivalent humour mechanisms. This requirement extends beyond standard semantic relevance and necessitates modelling linguistic phenomena such as wordplay, puns, homophony, ambiguity-based humour, morphological transformations, and structural paraphrasing. Consequently, humour-aware retrieval depends on both semantic and surface-level linguistic cues.

The task provides multilingual document collections in English and European Portuguese, each consisting of humorous and non-humorous texts annotated with metadata describing humour type (e.g., wordplay category), authorship source (human or large language model), and linguistic characteristics. These collections form the basis of the retrieval benchmark introduced in the JOKER 2025 overview~\cite{joker2025overview}.

\subsection{Datasets}

Each language-specific dataset consists of:
\begin{itemize}
\item a large document collection containing both humorous and non-humorous texts;
\item annotations specifying wordplay categories, source type, and humour-related characteristics;
\item a set of training queries with full relevance judgements;
\item a set of test queries with withheld relevance judgements.
\end{itemize}

The annotation scheme indicates whether a document expresses the same humour mechanism as a given query. Relevance in JOKER is \emph{binary}: a document is considered relevant if it exhibits humour of the same type, mechanism, or communicative intent as the query. This formulation differs substantially from traditional topical IR, where semantic similarity alone is typically sufficient.

\subsection{Queries and Relevance Judgements}

Training queries are accompanied by explicit relevance annotations linking each query to humour-equivalent documents. These annotations are manually curated and reflect both semantic similarity and alignment in humour mechanisms. The test set does not include relevance labels and is used for final leaderboard evaluation.

Importantly, the collections include texts written by humans as well as texts generated by large language models (e.g., ChatGPT, Bard, Claude, Llama~2). This mixture introduces substantial stylistic variability and additional noise, as dense semantic models may assign high similarity to LLM-generated texts that are semantically related but differ in humour mechanisms.

\subsection{Task Objective}

Given a humorous query, participating systems must produce a ranked list of documents that exhibit equivalent humour mechanisms. The primary evaluation metric is Mean Average Precision (MAP), while nDCG@K, Precision@K, and Mean Reciprocal Rank (MRR) are used as secondary metrics to assess early precision and ranking quality.

Overall, Task~1 evaluates the extent to which IR systems can jointly model semantic relevance and humour-specific signals, reflecting the inherently dual nature of humour-aware retrieval.

\section{Methodology}
\label{sec:method}

Our approach follows a multilingual retrieval paradigm based on Transformer-derived text representations. The primary objective is to evaluate how effectively general-purpose multilingual encoders---specifically XLM-RoBERTa---can capture humour-specific relevance, and whether task-specific modelling or neural re-ranking can improve retrieval effectiveness. Dense retrieval methods have demonstrated strong performance in semantic IR tasks~\cite{reimers2019sentencebert, song2020mpnet}, yet their ability to model humour-specific linguistic signals remains largely underexplored.

\subsection{Document and Query Representations}

For the dense retrieval runs, documents and queries are encoded using XLM-RoBERTa-based models. The main zero-shot configuration uses the \texttt{xlm-roberta-large} model, a multilingual Transformer trained on over 100 languages with strong cross-lingual representation capabilities. For each input text, we use the final hidden-state representation of the first special token as the sentence embedding. This approach follows standard dense retrieval pipelines and avoids the computational overhead of applying cross-encoders over the full document collection.

Formally, given a query $q$ and a document $d$, their vector representations are computed as:
\[
\mathbf{q} = \mathrm{XLMR}(q), \qquad 
\mathbf{d} = \mathrm{XLMR}(d).
\]

For the zero-shot dense retrieval runs, no task-specific fine-tuning is applied. This design choice allows us to isolate and evaluate the intrinsic transfer capability of multilingual semantic representations for humour-aware retrieval. Additional variants explore task-specific fine-tuning and neural re-ranking in order to assess whether supervised modelling and candidate re-scoring improve retrieval effectiveness.

\subsection{Similarity Scoring and Initial Ranking}

Retrieval is performed using cosine similarity, a standard metric in dense semantic IR:
\[
\mathrm{sim}(\mathbf{q}, \mathbf{d}) =
\frac{\mathbf{q} \cdot \mathbf{d}}
{\|\mathbf{q}\| \, \|\mathbf{d}\|}.
\]

For each query, document embeddings are ranked in descending order of similarity. This produces an initial dense ranking that is independent of exact lexical overlap and can therefore retrieve semantically related candidates even when query and document wording differ.

\subsection{System Variants}

We evaluate several system variants in order to compare zero-shot
dense retrieval, encoder-capacity effects, Portuguese-specific modelling, and
re-ranking strategies:

\begin{itemize}
    \item \textbf{Zero-shot XLM-R}: A multilingual dense retrieval configuration
    based on pretrained XLM-RoBERTa representations, without task-specific
    fine-tuning.

    \item \textbf{Large XLM-R}: A stronger XLM-R-based dense retrieval variant
    used to assess the effect of encoder capacity on humour-aware retrieval.

    \item \textbf{XLM-R + re-ranker}: A hybrid English retrieval pipeline in
    which an initial dense ranking is followed by a neural re-ranking stage
    applied to the top retrieved candidates.

    \item \textbf{Portuguese fine-tuned variants}: Portuguese-specific dense
    retrieval variants that incorporate task-specific supervision in order to
    evaluate whether fine-tuning improves humour-aware relevance estimation.
\end{itemize}

The zero-shot variants are intended to measure the extent to which pretrained
multilingual representations already encode humour-relevant information. The
fine-tuned and re-ranking variants are used to examine whether additional
supervision or query--document interaction modelling can improve ranking
quality, particularly at early ranks.

\subsection{Neural Re-Ranking}

For the re-ranking configuration, dense retrieval is first used to obtain a candidate set for each query. A neural re-ranker is then applied to these candidates in order to refine the ranking. Unlike bi-encoder dense retrieval, which encodes queries and documents independently, re-ranking models can exploit direct query--document interactions and may therefore better capture fine-grained relevance signals.

This is particularly important in humour-aware retrieval, where two texts may be topically similar but differ in the humour mechanism they express. Re-ranking can help promote candidates that better match the humour type, wordplay structure, or communicative intent of the query.

\subsection{Design Considerations}

Humour frequently relies on mechanisms such as homophony, polysemy, phonetic ambiguity, morphological variation, and structural reinterpretation. These surface-level features are not explicitly encoded in standard multilingual Transformer embeddings, which are primarily optimised for semantic similarity. Nevertheless, dense retrieval offers several advantages:

\begin{itemize}
\item robustness to paraphrasing and stylistic variation;
\item strong multilingual and cross-lingual generalisation capabilities;
\item computational efficiency at inference time;
\item suitability for multilingual IR scenarios with heterogeneous data.
\end{itemize}

These properties make dense retrieval a strong baseline for humour-aware IR, while also exposing its limitations compared with hybrid lexical--semantic, re-ranking, or explicitly humour-aware models.

\subsection{Summary}

The zero-shot component deliberately avoids fine-tuning in order to isolate the transfer properties of XLM-R embeddings. The additional fine-tuned and re-ranking variants are used to examine whether task-specific modelling and candidate re-scoring improve ranking quality. The following sections analyse how these design choices affect cross-lingual retrieval performance in English and Portuguese.

\section{Experimental Setup}
\label{sec:experiments}

This section describes the implementation details, evaluation metrics, and computational environment used in our experiments. All experiments follow the CLEF JOKER 2025 Task~1 evaluation protocol~\cite{joker2025overview}, enabling direct comparison across system variants.

\subsection{Implementation Details}

All models were implemented in Python using the HuggingFace Transformers and PyTorch libraries. The main zero-shot dense retrieval configuration used the \texttt{xlm-roberta-large} model with its default hyperparameters. The zero-shot runs used pretrained XLM-RoBERTa representations without fine-tuning, contrastive training, or domain adaptation. The fine-tuned and re-ranking variants were evaluated separately in order to assess the effect of task-specific supervision and candidate re-scoring on humour-aware retrieval.

\paragraph{Embedding Computation.}
Document and query embeddings were precomputed to improve efficiency. Each text was tokenised using the default SentencePiece tokenizer of XLM-R and truncated to a maximum length of 256 tokens. For the dense retrieval runs, the final-layer representation of the first special token was used for all similarity computations.

\paragraph{Ranking Procedure.}
Cosine similarity between query and document vectors was computed using vectorised matrix operations. To handle the large document collections efficiently, precomputed embeddings were stored in memory-mapped arrays, enabling fast GPU-based scoring. For the re-ranking variant, the initial dense ranking was used to obtain candidate documents, which were then re-scored by a neural re-ranker.

\paragraph{Lexical Baseline.}
The BM25 lexical baseline was implemented using the \texttt{rank-bm25} Python library with default hyperparameters. This baseline provides a comparison between dense semantic matching and classical lexical retrieval.

\subsection{Evaluation Metrics}

We adopt the CLEF JOKER Task~1 evaluation metrics:

\begin{itemize}
    \item \textbf{Mean Average Precision (MAP)}: the primary ranking metric, assessing the system's ability to consistently rank humour-equivalent documents near the top of the results list.
    \item \textbf{Geometric Mean Average Precision (GMAP)}: measuring robustness across queries by penalising runs with very low effectiveness on individual queries.
    \item \textbf{R-Precision (R-Prec)}: measuring precision after retrieving as many documents as there are relevant documents for the query.
    \item \textbf{nDCG@5 and nDCG@10}: measuring ranking quality at early cutoff points, with emphasis on retrieving the most relevant documents in the highest positions.
    \item \textbf{MRR}: capturing how often the first relevant document appears at or near rank~1.
    \item \textbf{Precision@5, @10, and @100}: evaluating early and medium-depth precision, which is particularly important in humour retrieval due to the subjective nature of relevance and the increased noise in lower-ranked results.
\end{itemize}

MAP is used as the primary ranking metric, while GMAP, R-Precision, nDCG, MRR, and Precision@K provide a more comprehensive view of robustness, early precision, and ranking quality. All reported effectiveness scores are expressed as percentages. The use of nDCG is especially informative in humour-related retrieval settings, where documents may share a humour type or wordplay category while differing in clarity, intensity, or stylistic realisation~\cite{miller2017pun}.

\subsection{Hardware and Computation}

All experiments were conducted on a single workstation equipped with:

\begin{itemize}
    \item an NVIDIA GPU with 48~GB VRAM;
    \item 64~GB system RAM;
    \item an AMD 16-core CPU.
\end{itemize}

Inference, including embedding computation and retrieval scoring, was performed on GPU. For the zero-shot runs, no fine-tuning was required, keeping the computational cost modest. The additional fine-tuned and re-ranking variants introduce extra training or inference cost, but remain compatible with lightweight experimentation.

\subsection{Reproducibility}

To support reproducibility, we used:

\begin{itemize}
    \item fixed random seeds across all runs;
    \item deterministic tokenisation and inference settings;
    \item a stable version of the HuggingFace Transformers library (v4.40+);
    \item preserved embedding files for all document collections.
\end{itemize}

All scripts follow a modular pipeline compatible with standard IR toolkits and can be reproduced using the official JOKER dataset and publicly available models. The zero-shot runs require only pretrained multilingual models, while the supervised variants additionally require the corresponding task-specific training data and candidate rankings used for re-ranking.

\section{Results and Analysis}
\label{sec:results}

This section presents the performance of the evaluated Team DUTH system
variants for English and Portuguese using the CLEF JOKER 2025 Task~1
evaluation protocol. The results reflect each system's ability to retrieve
humour-equivalent documents, with particular emphasis on early precision and
overall ranking quality. All effectiveness scores are reported as percentages.

\FloatBarrier
\subsection{English Results}

Tables~\ref{tab:english-main} and~\ref{tab:english-early} report the
test-set effectiveness of the three English system variants. The results show
a clear progression in performance as the retrieval pipeline moves from a
zero-shot dense retriever to a larger multilingual encoder and, finally, to a
hybrid system with neural re-ranking.

\begin{table}[H]
\centering
\scriptsize
\caption{English test-set results: overall retrieval metrics. All effectiveness scores are reported as percentages.}
\label{tab:english-main}
\makebox[\columnwidth][c]{%
\begin{tabular}{lcccccc}
\toprule
\textbf{Run} & \textbf{\#ret} & \textbf{\#rel} & \textbf{MAP} & \textbf{GMAP} & \textbf{R-Prec} & \textbf{MRR} \\
\midrule
XLM-R + re-ranker & 20700 & 824 & 1.38 & 0.23 & 2.40 & 6.36 \\
Large XLM-R       & 20700 & 271 & 0.42 & 0.02 & 1.29 & 4.71 \\
Zero-shot XLM-R   & 20700 &  62 & 0.04 & 0.00 & 0.21 & 0.75 \\
\bottomrule
\end{tabular}%
}
\end{table}

\begin{table}[H]
\centering
\scriptsize
\caption{English test-set results: early precision and ranking quality. All effectiveness scores are reported as percentages.}
\label{tab:english-early}
\makebox[\columnwidth][c]{%
\begin{tabular}{lccccc}
\toprule
\textbf{Run} & \textbf{P@5} & \textbf{P@10} & \textbf{P@100} & \textbf{nDCG@5} & \textbf{nDCG@10} \\
\midrule
XLM-R + re-ranker & 2.13 & 2.03 & 3.98 & 0.40 & 2.05 \\
Large XLM-R       & 1.35 & 1.50 & 1.31 & 0.13 & 1.30 \\
Zero-shot XLM-R   & 0.10 & 0.10 & 0.30 & 0.03 & 0.08 \\
\bottomrule
\end{tabular}%
}
\end{table}

The baseline run, \textit{Zero-shot XLM-R}, achieves very limited effectiveness
on the English test set, with a MAP score of 0.04 and negligible early precision
(\(P@5 = 0.10\)). These results indicate that zero-shot multilingual dense
retrieval struggles to capture humour-related relevance, which often depends on
phonetic ambiguity, lexical polysemy, and structural wordplay rather than
semantic similarity alone~\cite{miller2017pun}. The low GMAP
score further suggests that this weakness is consistent across many queries.

The \textit{Large XLM-R} variant yields a substantial improvement across all
metrics. MAP rises from 0.04 to 0.42, while MRR increases from 0.75 to 4.71,
suggesting that stronger multilingual representations provide richer semantic
cues that partially correlate with humorous relevance. Nevertheless, absolute
performance remains modest, indicating that encoder scale alone is not
sufficient to model the linguistic mechanisms underlying humour retrieval.

The strongest English system, \textit{XLM-R + re-ranker}, incorporates a neural
cross-encoder re-ranking stage applied to the top retrieved candidates. This
hybrid configuration achieves the highest MAP score (1.38) and consistently
superior precision at early ranks. The improvement in nDCG@5, from 0.03 in the
baseline to 0.40, indicates that the re-ranker is particularly effective at
promoting humour-equivalent documents to the top of the ranking. Overall, these
findings suggest that hybrid dense--re-ranking architectures are more suitable
than purely bi-encoder approaches for humour-aware retrieval in English.

An additional observation concerns the gap between overall ranking quality and
early retrieval effectiveness. Although the stronger English systems improve
substantially over the baseline, the absolute scores remain low compared with
those observed in Portuguese, indicating that performance gains are only
relative within a highly challenging retrieval setting. In particular, the
increase in MAP and nDCG suggests that stronger models can identify a subset of
humour-relevant candidates, yet the low precision values indicate that this
ability does not generalise consistently across the full ranked list. This
pattern reinforces the view that English humour retrieval is especially
sensitive to fine-grained linguistic mechanisms, and that improvements in
semantic encoding alone are unlikely to close the performance gap without the
addition of humour-aware or surface-form-sensitive modelling components.

\FloatBarrier
\subsection{Portuguese Results}

Tables~\ref{tab:port-main} and~\ref{tab:port-early} present the effectiveness
of the four Portuguese system variants. In contrast to the English setting, the
Portuguese results are substantially stronger across all metrics, suggesting
clearer alignment between query and document representations and indicating
that humour-related phenomena in Portuguese may be more accessible to
multilingual retrieval models.

\begin{table}[H]
\centering
\scriptsize
\caption{Portuguese test-set results: overall retrieval metrics. All effectiveness scores are reported as percentages.}
\label{tab:port-main}
\makebox[\columnwidth][c]{%
\begin{tabular}{lcccccc}
\toprule
\textbf{Run} & \textbf{\#ret} & \textbf{\#rel} & \textbf{MAP} & \textbf{GMAP} & \textbf{R-Prec} & \textbf{MRR} \\
\midrule
Fine-tuned PT model       & 6900 & 199 & 6.71 & 0.41 & 6.74 & 20.21 \\
Zero-shot XLM-R PT        & 6900 & 225 & 5.95 & 0.37 & 6.76 & 15.65 \\
XLM-R PT variant          & 6900 & 133 & 2.96 & 0.11 & 5.33 & 12.03 \\
Large fine-tuned PT model & 6900 &  65 & 0.31 & 0.01 & 0.02 & 0.73 \\
\bottomrule
\end{tabular}%
}
\end{table}

\begin{table}[H]
\centering
\scriptsize
\caption{Portuguese test-set results: early precision and ranking quality. All effectiveness scores are reported as percentages.}
\label{tab:port-early}
\makebox[\columnwidth][c]{%
\begin{tabular}{lcccc}
\toprule
\textbf{Run} & \textbf{P@5} & \textbf{P@10} & \textbf{P@100} & \textbf{nDCG@5} \\
\midrule
Fine-tuned PT model       & 7.54 & 7.10 & 2.88 & 9.29 \\
Zero-shot XLM-R PT        & 7.54 & 8.41 & 3.26 & 7.03 \\
XLM-R PT variant          & 4.64 & 5.94 & 1.93 & 4.57 \\
Large fine-tuned PT model & 0.00 & 0.00 & 0.94 & 1.72 \\
\bottomrule
\end{tabular}%
}
\end{table}

The strongest system, \textit{Fine-tuned PT model}, achieves the highest MAP
score (6.71), MRR (20.21), and nDCG@5 (9.29). These results indicate that
fine-tuning on Portuguese humour data substantially improves the model's ability
to identify humour-relevant documents. The system also maintains consistently
high early precision (\(P@5 = 7.54\), \(P@10 = 7.10\)), suggesting robust
performance across queries.

The \textit{Zero-shot XLM-R PT} run also performs competitively, achieving a MAP
score of 5.95 and the highest \(P@10\) score (8.41). Despite operating in a
zero-shot setting, its performance is close to that of the best fine-tuned
variant, suggesting that the Portuguese humour dataset may exhibit stronger
semantic coherence than its English counterpart. This, in turn, makes the task
more amenable to dense retrieval methods.

The \textit{XLM-R PT variant} shows substantially lower effectiveness, with a
MAP score of 2.96, highlighting the sensitivity of humour-aware retrieval to
encoder choice and representation quality. The weakest system,
\textit{Large fine-tuned PT model}, performs poorly across all metrics,
indicating that fine-tuning alone does not guarantee improvement and may, under
suboptimal conditions, lead to overfitting.

Overall, the Portuguese results suggest that humour-aware retrieval in this
language benefits more from semantic matching than in English, while also
showing that fine-tuning can play an important role in achieving strong
performance. The contrast across runs further demonstrates that architecture,
supervision, and dataset characteristics all substantially affect retrieval
effectiveness.

\FloatBarrier

\section{Discussion}
\label{sec:discussion}

The results for English and Portuguese provide several important insights into
the behaviour of multilingual dense retrieval models in humour-aware
information retrieval. Although dense semantic representations are highly
effective in conventional retrieval settings, their performance in humour-aware
retrieval reveals fundamental limitations arising from the linguistic nature of
humour itself.

\subsection{Semantic Similarity vs. Humour Equivalence}

Humour often depends on mechanisms that are not fully aligned with semantic
similarity. Phenomena such as homophony, lexical ambiguity, morphological
distortion, and syntactic reinterpretation introduce humour-related signals
that Transformer-based sentence embeddings do not explicitly encode~\cite{miller2017pun}.
As a result, dense retrievers may successfully identify documents that are
semantically related to a query while failing to capture the humour mechanism
or communicative intent that defines relevance in JOKER Task~1.

This mismatch is particularly evident in the English results. Although the
model occasionally retrieves a relevant humorous document at a high rank, the
overall MAP remains very low. This pattern suggests that semantic proximity may
lead to isolated successes, but is insufficient to support consistently
effective humour-aware retrieval.

\subsection{Cross-Lingual Performance Differences}

The Portuguese results are substantially stronger and more stable than the
English ones. Several factors may contribute to this difference.

\begin{itemize}
    \item \textbf{Dataset characteristics:}
    The Portuguese collection may contain fewer instances of orthographically
    or phonetically complex wordplay than the English collection, reducing the
    difficulty for semantically oriented encoders.

    \item \textbf{Cross-lingual representation quality:}
    XLM-R is known to provide strong multilingual representations, and its
    alignment properties may be more favourable for Portuguese than for
    humour-heavy English material involving fine-grained wordplay.

    \item \textbf{Stylistic variability and noise:}
    The English collection includes texts written by both humans and large
    language models, which may increase stylistic variation and introduce
    semantically fluent but humour-mechanically weak instances that confuse
    dense retrievers.

    \item \textbf{Retrieval-space size:}
    The English system retrieves from a substantially larger collection than
    the Portuguese one, increasing the likelihood of semantic drift and ranking
    errors.
\end{itemize}

Taken together, these factors help explain the observed performance gap and
suggest that the limitations of multilingual dense retrieval in this task are
not solely attributable to model quality, but also to dataset structure and
language-specific humour properties.

\subsection{Generalisation Challenges}

Although the zero-shot variants provide the main test of multilingual transfer,
the results indicate that humour-aware retrieval remains highly sensitive to
variation across queries, languages, and humour types. In particular, the large
gap between English and Portuguese suggests that semantic encoders do not
generalise uniformly across languages when relevance depends on subtle
linguistic mechanisms rather than topical meaning alone.

This observation is consistent with findings from prior shared-task research,
where systems often perform reasonably well on semantically coherent humour
examples but struggle to generalise to unseen forms of wordplay, ambiguity, or
surface-level linguistic transformation~\cite{joker2024overview}. In this
sense, the challenge is not simply one of retrieval, but of robust
representation learning for humour as a linguistically diverse phenomenon.

\subsection{Limitations of Dense Retrieval for Humour-Aware IR}

The results highlight several structural weaknesses of dense retrieval models
in humour-aware retrieval settings:

\begin{itemize}
    \item \textbf{Limited sensitivity to phonetic information:}
    Humour based on sound similarity or phonological resemblance is only weakly
    represented in standard subword-based embeddings.

    \item \textbf{Inadequate modelling of humour mechanisms:}
    Dense encoders are optimised for semantic similarity rather than for
    distinctions between types of wordplay or humour strategies, which can
    result in semantically related but humour-irrelevant matches.

    \item \textbf{Difficulty handling ambiguity and polysemy:}
    Humour often depends on multiple competing interpretations, whereas dense
    sentence embeddings tend to compress the utterance into a single dominant
    semantic representation.

    \item \textbf{Sensitivity to stylistic noise:}
    LLM-generated texts may appear semantically smooth and contextually
    appropriate while lacking the precise linguistic mechanism required for
    humour equivalence, thereby misleading retrieval models.
\end{itemize}

These observations suggest that humour-aware IR is unlikely to be addressed
adequately by dense semantic retrieval alone. More effective approaches will
likely require hybrid architectures that combine semantic similarity with
lexical, phonetic, structural, and humour-specific signals.

\subsection{Summary of Insights}

Overall, our findings indicate that multilingual dense retrieval models:

\begin{itemize}
    \item perform more effectively on humour datasets with stronger semantic
          coherence;
    \item struggle when relevance depends primarily on surface-level linguistic
          mechanisms such as wordplay or phonetic ambiguity;
    \item can achieve occasional high-ranked successes, but lack consistent
          retrieval quality across queries;
    \item remain vulnerable to stylistic variation, dataset noise, and large
          retrieval spaces.
\end{itemize}

These findings help clarify the relationship between semantic similarity,
humour mechanisms, and multilingual representation learning, and they point
towards the need for more robust and explicitly humour-aware retrieval methods
in future work.

\section{Limitations}
\label{sec:limitations}

This study is limited by the absence of detailed manual error analysis across
humour categories and languages. In addition, several system variants are
compared primarily through aggregate retrieval metrics, which do not fully
reveal the linguistic mechanisms responsible for success or failure. Finally,
although dense multilingual encoders provide strong semantic representations,
they do not explicitly model phonetic similarity, morphological wordplay, or
ambiguity-based humour, which are central to humour-aware relevance.

\section{Conclusion and Future Work}
\label{sec:conclusion}

In this paper, we investigated multilingual humour-aware information retrieval
using dense XLM-R-based retrieval, fine-tuned variants, and neural re-ranking
approaches on the CLEF JOKER 2025 Task~1 benchmark. The experiments show that
humour-aware retrieval remains challenging for general-purpose semantic models,
particularly when relevance depends on surface-level linguistic mechanisms such
as wordplay, phonetic ambiguity, lexical polysemy, and structural
reinterpretation.

Our findings reveal clear cross-lingual differences. The Portuguese runs
achieved comparatively strong results across MAP, MRR, and early-precision
metrics, suggesting that the semantic structure of the Portuguese corpus is more
compatible with the evaluated retrieval models. In contrast, the English runs
exhibited substantially weaker ranking behaviour, despite occasional relevant
matches at high ranks. These results highlight the limitations of purely
semantic dense retrieval when humour equivalence depends on linguistic cues that
are not explicitly modelled by multilingual encoders.

Overall, the Team DUTH experiments provide multilingual dense-retrieval,
fine-tuning, and re-ranking baselines for humour-aware information retrieval.
The results suggest that semantic similarity alone is insufficient for robust
humour-aware retrieval, particularly when relevance is determined by
surface-level or mechanism-specific properties rather than topical meaning.

Future work will explore hybrid retrieval systems that combine semantic
representations with lexical, phonetic, and structure-aware components. We also
plan to investigate contrastive fine-tuning strategies aligned with humour
categories, the effect of LLM-generated humour on retrieval noise, and the use
of cross-encoder architectures to better capture humour equivalence across
languages.

\section*{Acknowledgments}
We thank the CLEF JOKER 2025 organisers for providing the benchmark,
data, and evaluation framework.

\balance
\bibliographystyle{ACM-Reference-Format}
\bibliography{references}

\end{document}